\documentstyle[aps,twocolumn,psfig]{revtex}

\begin{document}

\twocolumn[\hsize\textwidth\columnwidth\hsize\csname @twocolumnfalse\endcsname

\title{Ferromagnetism in the Hubbard Model on the
Infinite Dimensional fcc Lattice}

\author{Martin Ulmke}

\address{Department of Physics, University of California, Davis,
CA 95616, USA}

\date{\today}

\maketitle

\begin{abstract}

The Hubbard model on the fcc lattice is studied in the limit of
infinite spatial dimensions. At sufficiently strong interaction
finite temperature Quantum Monte Carlo calculations yield a second
order phase transition to a highly polarized ferromagnetic state.
Ferromagnetism is stable over a wide hole doping regime but not for
electron doping. A possible non-Fermi liquid behavior is discussed.
\\

\\

\end{abstract}

]

More than 30 years ago, the Hubbard model was introduced to
describe correlation effects in transition metals, in particular the
band ferromagnetism of Fe, Co and Ni. However, it appeared to be a generic
model rather for \em anti-\em ferromagnetism than for ferromagnetism.
In fact, our knowledge on the possibility of ferromagnetism in the
Hubbard model is still very limited.
Only in a few special cases was the Hubbard model proven to have
ferromagnetic order. The first rigorous result showing a fully polarized
ferromagnetic ground state, the theorem by Nagaoka and Thouless \cite{1,2},
is only valid for a single hole added to the half filled band in the limit
of infinite on-site repulsion. Its thermodynamic relevance is unclear
until today. In the opposite limit of low electronic density a fully polarized
ferromagnetic ground state has been found recently by M\"uller-Hartmann
\cite{3} in the Hubbard model on a zigzag chain in the continuum limit.
Further, the existence of a ground state with net polarization has been
proven for the half filled band case on bipartite lattices with asymmetry
in the number of sites per sublattice \cite{4}, and in 'flat-band' systems
which have a ferromagnetic ground state for any interaction strength
$U$ larger than zero \cite{5,6}.

Since ferromagnetism in the Hubbard model has been found only in some
particular cases one might suspect that the model is just too
oversimplified to explain band ferromagnetism even in a qualitative way.
One missing feature relevant in real materials is band degeneracy.
A mechanism similar to Hund's rule in individual atoms could lead to a
ferromagnetic ordering of the electronic spins.
In addition, all non-local interactions are neglected in the Hubbard model.
The question if nearest-neighbor interactions, in particular the Coulomb
exchange and a pair hopping term, are able to induce ferromagnetism
has been adressed by Hirsch and Tang \cite{7,8}. Within mean-field
approximations and Monte Carlo simulations in one dimensions at half filling
they found fully and partially polarized ground states.
Recently, Strack and Vollhardt derived rigorous criteria for the stability
of saturated ferromagnetism in an extendend single band Hubbard model
including all nearest neighbor interactions at half filling \cite{9,10}.
Again, the exchange interaction plays an important role in stabilizing
ferromagnetism.
%The criteria could be generalized to general two-site
%interactions \cite{10} and a magnetic field in the pure model \cite{11}.

However, since the results are valid only for the insulating half filled
band case, the question about the necessary microscopic ingredients
for metallic ferromagnetism remains open.

In this paper we are concerned with the 'pure' Hubbard model, i.e.
a single electronic band with only on-site interaction. The question
we want to answer is:\\

Is the pure Hubbard model capable of describing metallic ferromagnetism?\\

Within the pure Hubbard model we still have freedom in the choice of
the lattice geometry. We choose the fcc lattice for the following reasons:
i) It is not bipartite. Thus we expect the competing antiferromagnetic
order to be suppressed or at least weakend by frustration.
ii) The fcc lattice has been shown to be a good 'environment' for
ferromagnetism by variational calculations of the stability of the
Nagaoka state against
doping \cite{11,12,13}, by exact diagonalization studies \cite{13,14}, and also
by nature since Ni with fcc structure is fully polarized whereas Fe
(bcc-structure) is not, as pointed out in ref.~\cite{12}.
Since the Hubbard model cannot be solved in finite dimensions, $d>1$, and
Monte Carlo simulations would allow only very small fcc lattices we study it
in the limit of infinite dimensions, $d=\infty$ \cite{15}. Here, the
system is reduced to a dynamical single site problem \cite{16} equivalent
to an Anderson impurity model complemented by a self-consistency condition
\cite{17}. Still, it cannot be solved analytically, and to avoid further
approximations we employ a finite temperature Quantum Monte Carlo method
\cite{18}. This approach has proven to be a powerful and reliable tool
for the study of correlated fermi systems both for thermodynamical
as well as dynamical properties \cite{19}.
Spectral properties are obtained by analytic continuation of the imaginary
time data using the Maximum Entropy method \cite{20}.

A non-trivial generalization of the fcc lattice in $d$ dimensions \cite{21}
is the set of all points with integer cubic coordinates summing up to an even
integer. It is a non-bipartite Bravais lattice for any dimension $d>2$.
Each point has $Z=2d(d-1)$ nearest neighbors defined by all vectors $\bf x$
which can be written in the form $\bf x = \pm \bf{e}_i \pm \bf{e}_j$,
with two different cubic unit vectors
$\bf{e}_i$ and $\bf{e}_j$ ($i,j=1,\cdots,d$).
With the proper scaling of the hopping term, $t=1/\sqrt{Z}$ \cite{15},
the density of states of the non-interacting electrons can be calculated
in $d=\infty$ \cite{21} to be:
\begin{equation}
N_0(E) = e^{-(1+\sqrt{2}E)/2}/\sqrt{\pi (1+\sqrt{2}E)}
\label{Gl1}
\end{equation}
where we choose contrary to usual convention a \em positive \em hopping
integral which gives a strong square-root divergency at the lower band edge,
$-1/\sqrt{2}$, and no upper band edge \cite{22}.
The lattice structure enters in the $d=\infty$-approach only via the density
of states. The non-bipartite structure of the fcc lattice leads to the
strong asymmetry in $N_0(E)$.

The variational treatment of the Nagaoka state on the fcc lattice
\cite{11,12,13}
provides upper limits for the critical doping $\delta_c$ and lower limits for
the critical interaction $U_c$ where saturated ferromagnetism becomes unstable.
While for the simple cubic lattice in $d=\infty$ the stability regime
shrinks to the point $U_c=\infty$, $\delta_c=0$ \cite{23},
for the fcc-density of states (1) the Nagaoka state is stable above
a critical line $U_c(n)$ with $U_c(0)=0$ and $U_c(1)=\infty$ \cite{24}.
The Nagaoka state is unstable in the case of electron doping $(n>1)$.
Antiferromagnetism is not expected on the fcc-lattice in high dimensions
because the difference of the numbers of not frustrated bonds and frustrated
bonds is only of the order of $d$  resulting in an effective field of the
order of $t^2 d\propto 1/d$ \cite{24}.

To detect a ferromagnetic instability we calculate the corresponding
susceptibility, $\chi_F$, from the two-particle correlation functions
\cite{25}. We choose a relatively strong interaction
of $U=4$ \cite{26} and an electronic density of $n=0.58$, where the
ferromagnetic response appeared to be strong. As a function of temperature,
we find $\chi_F$ to obey a Curie-Weiss law and extrapolate the transition
temperature linearly from the zero of $\chi_F^{-1}$ (see Fig.~1).
The numerical result is a Curie temperature of $T_c=0.051(2)$ \cite{27}.
We also calculate the magnetization, $M$, within the ferromagnetic phase,
i.e.~below $T_c$. $M$ grows strongly at lower temperatures reaching more
than 80\% of the fully polarized value ($M_{max}=n=0.58$) at the lowest
temperature, $T=0.0357$. A fit through the three data points with a Brillouin
function (Fig.~1) would give a full polarization at $T=0$ and the same
critical temperature of $T_c=0.05$ as above.
Translated to the three dimensional fcc lattice with $Z=12$ nearest
neighbors and a bandwidth of $W=16t$ the critical temperature becomes
$T_c(3D)\approx 0.011 W$.
Thus, despite the oversimplifications of the single band Hubbard model,
the resulting Curie temperature has a realistic order of magnitude of
500-800K for typical values of $W$ around 5 eV.

In order to answer the question if the system is still metallic
we calculate the single particle spectrum. In the ferromagnetic
phase we can distinguish two different spectra for spins parallel
and anti-parallel to the net magnetization (Fig.~2).
The latter (minority spin) spectrum (dotted line) is shifted
to higher frequencies and has a pronounced upper band around
$\omega-\mu\approx U=4$ with a developing (pseudo) gap.
The majority spin spectrum, however, is only slightly affected by the
interaction. Here, the weight of the upper band is very small since due
to the Pauli principle the majority spin-electrons are unlikely to hop
over preoccupied sites.
Apparently the system is metallic since both spectra have a finite value
at the Fermi level ($\omega=\mu$). This also holds in the paramagnetic phase
(not shown).
\vskip-25mm
\begin{figure}
\psfig{file=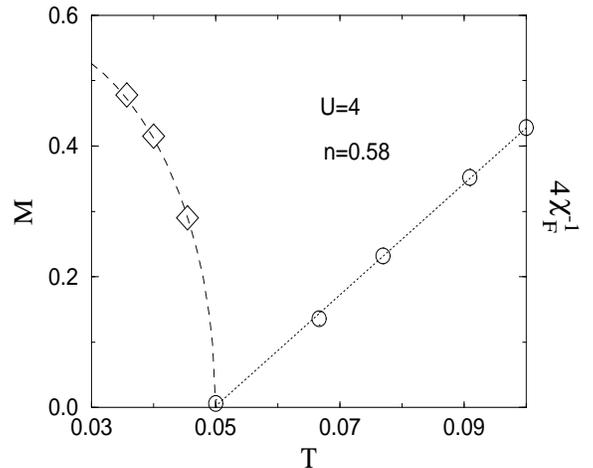,height=4.3in,width=5.6in}
\vskip-15mm
\caption{Magnetization, $M$ (diamonds), and inverse ferromagnetic
susceptibility, $\chi_F^{-1}$ (circles; values multiplied by a factor of 4
to use the same scale) for $U=4$ and $n=0.58$. Error-bars are of the size of
the symbols or smaller. Dotted line is a linear fit to $\chi_F^{-1}$,
dashed line a fit with a Brillouin function to $M$.}
\end{figure}
\vskip-30mm
\begin{figure}
\psfig{file=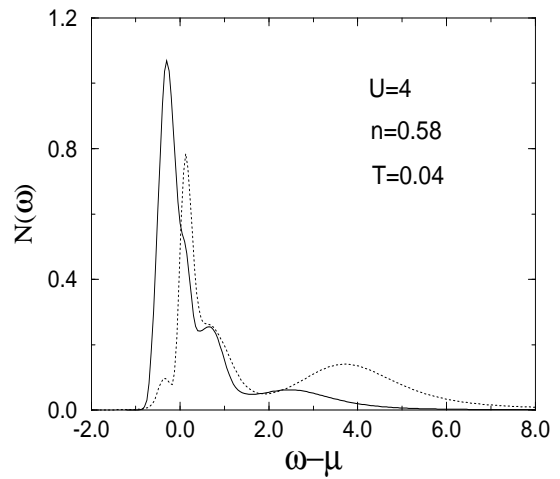,height=4.3in,width=5.6in}
\vskip-15mm
\caption{ Single particle spectrum, $N(\omega)$, for $U=4$, $n=0.58$,
and $T=0.04$. Full (dotted) line: majority (minority) spin spectrum.
}
\end{figure}
Seemingly our finding violates the general theorem on local Fermi liquids
(LFL) \cite{28} that prohibits ferromagnetic instabilities in a single band
Fermi liquid with local self energy. How can this contradiction be resolved?
In the present situation the self energy $\Sigma_n\equiv\Sigma(i\omega_n)$
is local by construction, therefore we investigate its imaginary part to
determine if the system is a Fermi-liquid. In that case the imaginary
part of $\Sigma_n$ vanishes at least linearly with Matsubara frequency
$|\omega_n|\to 0$.
Although our finite temperature approach does not allow to reach the limit
$\omega_n \to 0$, the results close above and below the transition
(Fig.~3) strongly suggest that Im $\Sigma_n$ approaches a finite
value at low frequencies or at least vanishes much slower than $\omega_n$.
Together with results for the single particle spectrum we conclude
that the system is likely to be a metal with non-Fermi liquid properties.
The existence of a metallic non-Fermi liquid state in high dimensions
is quite surprising but has been observed before in an extended Hubbard
model \cite{29} and also in an alloy analogy approach to the single band
Hubbard model over a wide range of doping and interaction strength \cite{30}.
It can also not be ruled out by existing 'essentially exact' treatments
of the Hubbard model on infinite dimensional bipartite lattices \cite{19}.
Further investigations on the nature of the paramagnetic phase are required
to settle this question.
\vskip-25mm
\begin{figure}
\psfig{file=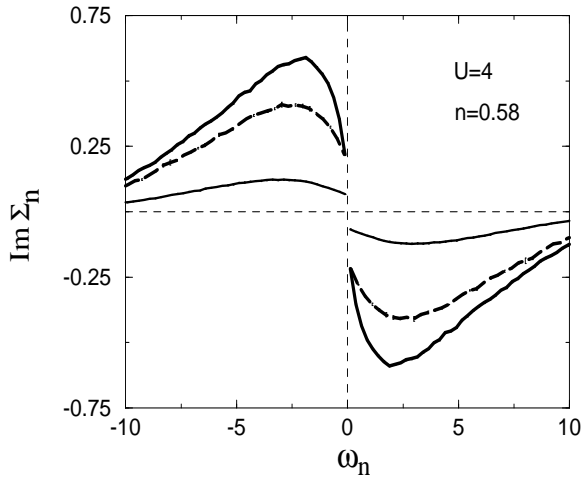,height=4.3in,width=5.6in}
\vskip-15mm
\caption{Imaginary part of the self energy for $U=4$ and $n=0.58$ in the
paramagnetic phase ($T=1/13, 1/15, 1/20$, dashed line; curves are on top of
each other) and in the ferromagnetic phase ($T=1/25$, solid lines;
smaller absolute values correspond to the majority spin).
Statistical errors are within the width of the lines.
}
\end{figure}
Finally, we investigate the doping dependence of the critical temperature,
$T_c$ (Fig.~4). We find positive values of $T_c$ only in the case of hole
doping and there over almost the whole density regime.
Only above a density of $n\approx 0.9$ the extrapolation described above
would lead to negative values of $T_c$, but with the present approach we
cannot rule out very low transition temperatures of the order of $10^{-3}$.
For low densities the Curie temperature becomes very small and seems to
vanish close to $n=0$. Within the error-bars the results are consistent with
the variational approach which yields a $T_c$ proportional to $n^3$
for small $n$, i.e.~no lower critical electron density \cite{24}.
We also checked for phase separation by calculating the electronic
compressibility but never observed any instability.

The phase diagram in Fig.~4 is qualitatively similar to the results of a
'self consistent moment approach' to the the Hubbard model on the
three dimensional fcc-lattice \cite{31}. In particular the optimal electron
density is found to be $n\approx 0.65$ which is consistent with our results.
\vskip-25mm
\begin{figure}
\psfig{file=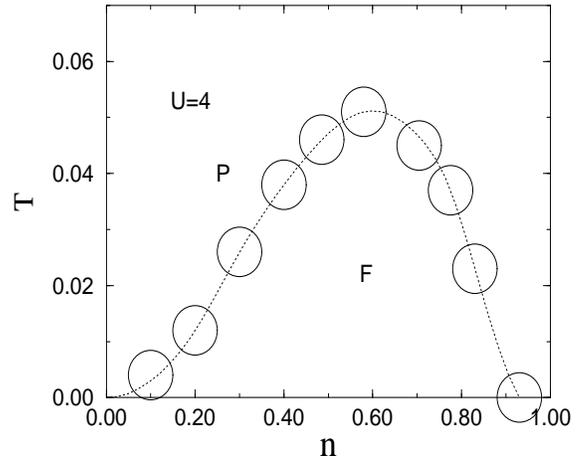,height=4.3in,width=5.6in}
\vskip-15mm
\caption{Density dependence of the Curie temperature $T_c$.
P and F denote para- and ferromagnetic phases.
Dotted line is a guide to the eye only.}
\end{figure}
In summary, we presented a numerical proof of metallic ferromagnetism
in the single band Hubbard model in the limit of infinite dimension.
An asymmetric density of states, corresponding to a non-bipartite lattice,
appears to be an essential ingredient for ferromagnetism.
For a given interaction strength ferromagnetism is found over a wide
hole doping regime but not for electron doping.

The author acknowledges helpful discussions and correspondence with K.~Held,
E.~M\"uller-Hartmann, V.~Jani\v{s}, A.~Sandvik, R.~Scalettar, D.~Vollhardt
and G.~Zimanyi.
He thanks A.~Sandvik for kindly providing his Maximum Entropy program.

This work was supported by a grant from the Office of Naval Research,
ONR N00014--93--1--0495 and by the Deutsche Forschungsgemeinschaft.


\begin{references}
\bibitem{27} The data in Fig.~1 are obtained for one small but fixed
discretization $\Delta\tau=0.25$ in imaginary time. The error in $T_c$ is
predominantly due to the additional extrapolation to $\Delta\tau\to 0$.
\end{references}
\end{document}